\documentclass{pasj00}

\usepackage{txfonts}

\begin{document}

\SetRunningHead{}{}
\Received{}
\Accepted{}

\title{A Discovery of a Peculiar Pulsar in the Small Magellanic Cloud}

\author{Masaru \textsc{UENO}, Hiroya \textsc{YAMAGUCHI}, 
Shin-ichiro \textsc{TAKAGI}, Jun \textsc{YOKOGAWA},
and Katsuji \textsc{KOYAMA}}
\affil{Department of Physics, Graduate School of Science, Kyoto University, 
Sakyo-ku, Kyoto\ 606-8502}
\email{masaru@cr.scphys.kyoto-u.ac.jp}

\KeyWords{Pulsars: individual (IKT~1) --- Stars: neutron 
--- X-rays: spectra --- X-rays: transient}

\maketitle

\begin{abstract}

We report on a peculiar X-ray binary pulsar IKT~1 = RX~J0047.3$-$7312  
observed with XMM-Newton in Oct.\ 2000.
The X-ray spectrum is described by a two-component spectrum.
The hard component has a broken power-law with respective photon 
indices of 0.2 and 1.8, below and above the break energy
at 5.8~keV. The soft component can be modeled by 
a blackbody of $kT = 0.6$~keV.
The X-ray flux shows a gradual decrease and periodic variations of 
about 4000 s. The averaged flux in 0.7$-$10.0~keV is 
$2.9 \times 10^{-12}$~ergs~cm$^{-2}$~s$^{-1}$, which is $\sim 10$ times 
brighter than that in a ROSAT observation in Nov.\ 1999. 
In addition to the 4000-s variation, we found coherent pulsations 
of 263 $\pm$ 1~s. These  discoveries strengthen 
the Be/X-ray binary scenario proposed 
by the ROSAT and ASCA observations on this source, and 
confirm that most of the hard sources in the Small Magellanic Cloud 
are X-ray binary pulsars. 
A peculiar property of this XBP is that the coherent pulsations are 
found only in the soft component, and the folded light curve 
shows a flat top shape with a sharp dip.
We discuss the nature of this XBP focusing on the peculiar soft 
component.

\end{abstract}

\section{Introduction}

More than 30 X-ray binary pulsars (XBP) have been discovered 
in the Small Magellanic Cloud 
(SMC) by the recent satellites such as 
ASCA, RXTE, Chandra, and XMM-Newton (e.g., \cite{yokogawa2003}). 
Most of the XBPs in the SMC have been  classified, or proposed
to be Be/X-ray binaries with the  estimated  ages
of $\sim 10^7$~yrs. 
The number of the XBPs normalized to the galaxy mass is 
far larger in the SMC than in our Galaxy (\cite{haberl2000}; 
\cite{yokogawa2003}). This high population of XBPs would be
good evidence for the active star formation in the SMC 
of $\sim 10^{7}$~yrs ago. 

Other than the population study of the X-ray sources, 
the SMC is a good local galaxy for the study of
individual X-ray sources such as the emission mechanism of XBPs, 
because the interstellar extinction 
for the member sources is very low and the line-of-sight
galaxy depth is small ($\sim$ 10~kpc) compared
to the SMC distance of $\sim 60$~kpc 
(\citet{harries} and references therein). 
The former gives us soft X-ray information down to a few-hundred eV,
while the latter provides us with a more accurate distance, and hence 
the X-ray luminosity, than those of the galactic sources. 
The large collective area of XMM-Newton enables us to
perform  the X-ray spectroscopy down to the 
luminosities of $\sim 10^{34}$~ergs~s$^{-1}$ (e.g., \cite{sasaki2003})
even for  more distant (60~kpc) XBPs than those in our 
Galaxy (typically 1$-$20~kpc).

IKT~1 was first discovered by \citet{inoue} with 
Einstein. ROSAT observed this source many times and 
\citet{haberl2000} proposed this source to be a Be/X-ray 
binary candidate from the flux variability and the existence of 
an optical counterpart with emission lines. ASCA also detected 
this source and found a flare-like behavior (\cite{yokogawa2003}). 
\citet{yokogawa2003} plotted hardness ratios of 
X-ray sources in the SMC against their luminosities 
and found that XBPs reside in a restricted region. 
Since IKT~1 is located in this region, IKT~1 is likely to be an XBP. 

Using the XMM-Newton archive data, we found coherent X-ray pulsations 
from this source. This paper reports on the discovery of a 263-s pulsation 
and the timing and spectrum analysis. 
Discussion on the nature of IKT~1 is made focusing on 
the phase-resolved spectroscopy. 
Independent detection of the 263-s pulsation was recently 
reported by Haberl \& Pietsch\footnote{Haberl \& Pietsch 
appeared in astro-ph (0311092, accepted to A\&A) after 
we submitted the present paper.}.
The source distance is assumed to be 60~kpc, 
the same value as the SMC \citep{harries}. 

\section{Observation}

The main objective of the XMM-Newton (\cite{aschenbach}) 
observation made on Oct.\ 15, 2000 was to study a 
supernova remnant (SNR), IKT~5 \citep{inoue}, 
and other X-ray sources around this SNR.
In this field, we found IKT~1 with
a higher X-ray flux  than the previous observations of ROSAT and ASCA,
and hence we study the X-ray properties of this source.
Since EPIC/PN \citep{struder} has better
time resolution and statistics  than MOS1/2 \citep{turner},
and IKT~1 accidentally fell in the CCD dip of MOS2, we concentrate
on the X-ray data from EPIC/PN, and use the MOS1 and 2 data 
for the consistency check. 

PN was operated in the extended full frame mode with 
the medium filter and the time resolution of 200~ms. 
We  used the version 5.4.1 of 
the Standard Analysis System (SAS) software 
for the event selection. From the detected PN events, 
we selected those with the PATTERN keywords between 0 and 4
as X-ray events. The exposure time was $\sim$ 22~ks. 

\section{Results}

\subsection{Image Analysis}

\begin{figure}
   \begin{center}
      \FigureFile(80mm,50mm){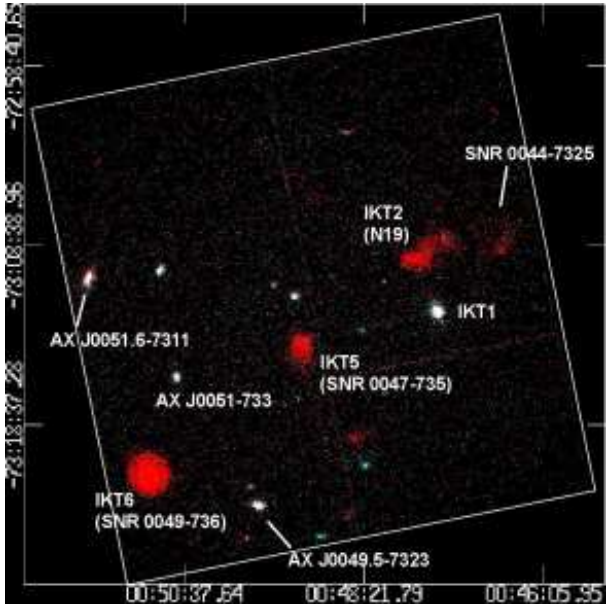}
   \end{center}
   \caption{XMM-Newton/PN image around IKT~5. The soft (0.5$-$2.0~keV) and 
hard (2.0$-$7.0~keV) band images are represented with red and blue colors, 
respectively. The images have been smoothed to a resolution of 5''. 
XBPs and SNRs in the FOV are designated with their names. 
The FOV of the PN detector is shown with the solid square. }
\label{fig:image}
\end{figure}

The hard (2.0$-$7.0~keV) and soft (0.5$-$2.0 keV) band
X-ray images are shown with two colors in Figure~\ref{fig:image}. 
Three XBPs (AX~J0049.5$-$7323, AX~J0051$-$733, AX~J0051.6$-$7311; 
\cite{yokogawa2000c}c, \cite{imanishi}, \cite{yokogawa2000b}b) 
and 4 SNRs (0044$-$7325, IKT~2 = N19, IKT~5 = 0047$-$735, 
IKT~6 = 0049$-$736; e.g., \cite{wang}) 
are clearly found  in the field of view (FOV). 
All the XBPs appear as white point sources, 
and the SNRs show red extended structures. 
Blue color sources may be extra galactic AGNs.
All the three pulsars mentioned above are identified with 
emission line objects (ELOs) cataloged by \citet{meyssonnier} 
\citep{haberl2000}. 
The coordinates of these pulsars derived from the PN data  
are systematically shifted from  the positions of the optical counterparts
by ($\Delta$R.A., $\Delta$Dec.) = ($-$\timeform{2.4''}, $-$\timeform{0.1''}).
We hence fine-tuned the PN coordinates to the optical coordinates. 
After this tuning, the root-mean-square of the differences 
between the PN and the optical positions for 
the three pulsars is ($\Delta$R.A., $\Delta$Dec.) = (\timeform{0.2''}, 
\timeform{0.7''}). This can be used as a typical positional error
between the two frame. 

The brightest source in the FOV is the white point source 
at the south of IKT~2. The fine-tuned  position is 
R.A. = \timeform{00h47d23s.3}, Dec. = $-$\timeform{73D12'27''} (J2000), 
and hence we designate this source as XMMU~J004723.3$-$731227. 
Although $1\sigma$-statistical error of this source position
is only \timeform{0.1''}, a realistic error would be 
$\sim$ \timeform{0.7''} (see the previous paragraph).
Since all the error regions of  XMMU~J004723.3$-$731227, 
IKT~1 \citep{inoue}, and RX~J0047.3$-$7312 
\citep{haberl2000} overlap, we conclude 
these three sources are the same X-ray object.

\subsection{Timing Analysis}

\begin{figure}
   \begin{center}
      \FigureFile(80mm,50mm){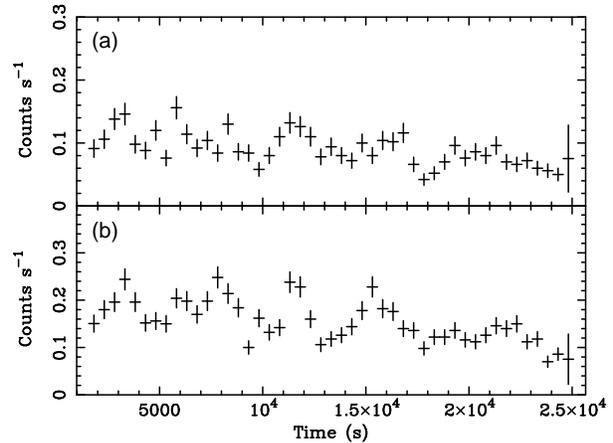}
   \end{center}
   \caption{Light curves of IKT~1 in the 0.5$-$2.0 (a), and 
2.0$-$10.0~keV (b) bands.}
\label{fig:lc}
\end{figure}

For the timing analysis of IKT~1, 
X-ray photons were extracted from a circular 
region of 40'' radius centered on IKT~1 in the PN image. 
The light curves with the time bin of 500~s 
in the energy bands of 0.5$-$2.0 and 
2.0$-$10.0~keV are shown in Figure~\ref{fig:lc}.
In both the energy bands, we see several flare-like events
on the general trend of decreasing flux. 

\begin{figure}
   \begin{center}
      \FigureFile(80mm,50mm){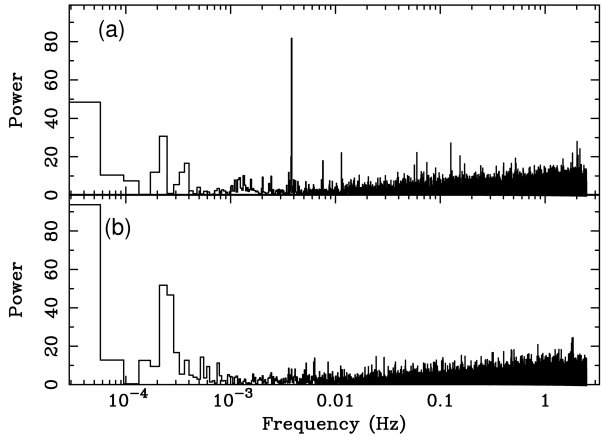}
   \end{center}
   \caption{Power spectra of IKT~1  in 
the 0.5$-$2.0 (a) and 2.0$-$10.0~keV (b) bands. 
The power is normalized to be 2 for random fluctuations. 
The peak at $\sim 3.82 \times 10^{-3}$~Hz is 
seen only in the 0.5$-$2.0~keV band.}
\label{fig:psd}
\end{figure}

After the barycentric arrival time corrections, 
we searched for periodicity in the energy bands 
of 0.5$-$2.0 and 2.0$-$10.0~keV, using a Fast Fourier 
Transformation algorism. Figure~\ref{fig:psd} 
shows the resultant power density spectra 
in the 3.8$\times 10^{-5}-$2.5~Hz frequency band. 
In the 0.5$-$2.0~keV band, a maximum power of 81.6 was 
obtained at the frequency of 
$3.82\times 10^{-3}$~Hz ($=262$~s). 
Since the probability to detect such a large power 
in any frequency from random events is only $\sim 1\times 10^{-13}$, 
the detection of the coherent pulsations is highly significant. 
In the 2.0$-$10 keV band, however, we see no significant power excess
at this frequency.
The other large power peak at $2.29 \times 10^{-4}$~Hz is
due to the periodic flares.  Unlike the coherent pulsations, this peak is  
found in both the energy  bands. 

\begin{figure}
   \begin{center}
      \FigureFile(80mm,50mm){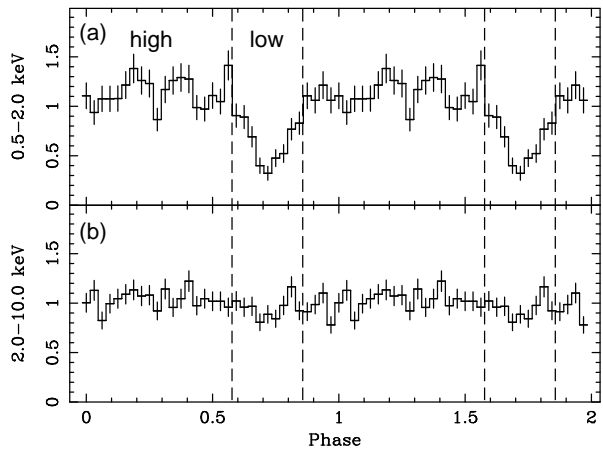}
   \end{center}
   \caption{Pulse profiles folded with the period of 
263~s in the two energy bands: 0.5$-$2.0~keV (a), and 
2.0$-$10.0~keV (b). The phase averaged count rates 
are normalized to be unity.}
\label{fig:fold}
\end{figure}

We next  performed epoch folding searches for 
the periods near the two power-peak frequencies. 
The flare-like events are found to have the most probable period 
of $3920 \pm 150$~s. The barycentric period of 
the coherent pulsations is determined to be 
$P=263 \pm 1$~s. Figure~\ref{fig:fold} 
shows the folded pulse profiles at the period of 263~s 
in the 0.5$-$2.0 and 2.0$-$10.0~keV bands. 
We see no pulse modulation in the hard X-ray band, 
while the pulse profile found in the soft band 
is very peculiar: a flat top profile with a V-shape dip.

\subsection{Spectral Analysis}

The X-ray spectrum was extracted from the same region 
as the timing analysis (the 40''-radius circle), 
while the background spectrum was extracted 
from the annular region around the source 
with the inner and outer radii of 40'' and 80'', respectively. 

\begin{figure}
   \begin{center}
      \FigureFile(80mm,50mm){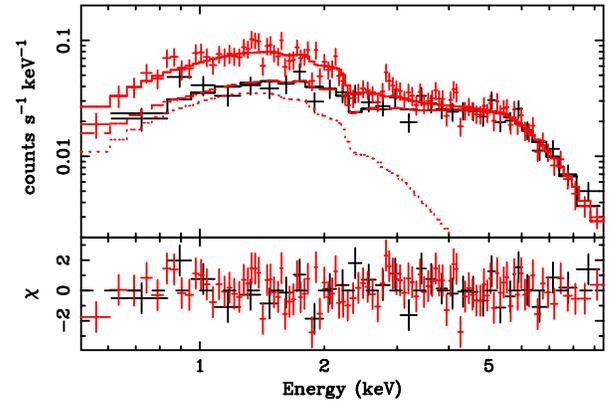}
   \end{center}
   \caption{(Upper panel:) Background-subtracted 
spectra in the ``low'' (black) and ``high'' (red) phases. 
The best-fit models and the blackbody component 
is shown with the solid and dot lines, respectively.
(Bottom panel:) Residuals of the source counts from 
the best-fit models.}
\label{fig:spec}
\end{figure}

Since the pulse profiles are energy dependent, we made
X-ray spectra separately, during the pulse minimum (low) 
and maximum (high) phases as shown in Figure~\ref{fig:fold}.
The background subtracted spectra for the low and high phases 
are given in Figure~\ref{fig:spec}. 
In the hard X-ray band above $\sim$3 keV, the two X-ray spectra 
show essentially the same profile and flux, 
while below $\sim$2 keV the spectrum of the high phase shows 
a large excess over that of the low phase. 
This is in excellent agreement with that the X-ray pulsation
is found only in the soft X-ray band.

Since both the spectra have a break at $\sim$6~keV, 
we fitted the spectra with a broken power-law model. 
The photon indices, break energies and normalizations were 
simultaneously fitted, while the absorption columns were 
treated independently for each spectrum. 
In this paper, the photoelectric absorption was 
calculated using the cross sections by \citet{morrison1983} 
assuming the solar abundance ratio \citep{anders1989}. 
This model reproduced the hard energy band well, 
but left large residuals at the soft energy band, 
and was statistically rejected 
with a $\chi^{2}/$degree of freedom (d.o.f.) = 160.9/129. 
Therefore the X-ray pulsation is not due 
to a periodic variation of the absorption. 

We then added a blackbody (BB) as the soft component 
to the broken power-law model and performed 
a simultaneous fit for the low- and high-phase spectra. 
The normalizations of the BB component 
were treated as independent parameters between the two 
phases. As a result, the fit was acceptable 
with a $\chi^{2}$/d.o.f.=125.9/127, 
with the best-fit parameters and models shown 
in Table \ref{tab:spec} and Figure~\ref{fig:spec}, respectively. 

\begin{table}
\begin{center}
\caption{Best-fit parameters with a broken power-law plus blackbody model.}
\label{tab:spec}
  \begin{tabular}{lc}
   \hline \hline
   Parameters & Value\\
  \hline
   Broken power-law  & \\   
   \hspace*{0.3cm}$\Gamma _{1}$\footnotemark[$*$]\dotfill  & $0.19_{-0.30}^{+0.09}$\\
   \hspace*{0.3cm}$\Gamma _{2}$\footnotemark[$*$]\dotfill  & $1.8_{-0.4}^{+0.6}$\\
   \hspace*{0.3cm}Break Energy\dotfill   & $5.5_{-0.6}^{+0.4}$\\
   \hspace*{0.3cm}Norm\footnotemark[$\dagger$]\dotfill     & $6.7_{-2.0}^{+1.0}$\\
   Blackbody \hfill &\\
   \hspace*{0.3cm}$kT$~(keV)\dotfill     & $0.59_{-0.10}^{+0.11}$\\
   \hspace*{0.3cm}$L_{\rm total}$ (``low'')\footnotemark[$\ddagger$]\dotfill &$<0.5$\\
   \hspace*{0.3cm}$L_{\rm total}$ (``high'')\footnotemark[$\ddagger$]\dotfill& $1.4_{-0.3}^{+0.6}$\\
   $N_{\rm H}$  [cm$^{-2}$]\dotfill  & ($9.6_{-4.4}^{+4.6}) \times 10^{20}$\\
\hline
  \multicolumn{2}{@{}l@{}}{\hbox to 0pt{\parbox{60mm}{\footnotesize
      Notes. Error regions correspond to 90\% confidence levels. 
      \par\noindent
      \footnotemark[$*$]Photon indices below ($\Gamma _{1}$) and above ($\Gamma _{2}$) the
      break energy. 
      \par\noindent
      \footnotemark[$\dagger$]Normalization at 1 keV ($\times 10^{-5}$~photons~keV$^{-1}$~cm$^{-2}$~s$^{-1}$). 
      \par\noindent
      \footnotemark[$\ddagger$]Total luminosity 
      ($\times 10^{35}$~ergs~s$^{-1}$) of the blackbody component 
      in the ``low'' and ``high'' phases at the SMC distance 
      (60~kpc; \cite{harries}). 
    }\hss}}
  \end{tabular}
\end{center}
\end{table}

As for  the 4000-s periodicity, we also made 
separate spectra for the  high  and low flux phases 
using the 4000-s folded light curve. These spectra are 
simultaneously fitted with the same two-component model as 
that for the  case of the coherent pulsation (263~s); 
we fixed the spectral parameters to the best-fit values 
given in Table \ref{tab:spec} except for the normalizations.
In this case, we found an acceptable fit when the BB component 
is constant but only the normalizations of 
the broken power-law component varied between 
the high and low phases. Therefore the 4000-s variation 
is attributable to the broken power-law component.

\section{Long-term Flux History}

To investigate a long-term flux variability,
we accessed the HEASARC archive and found that IKT~1 was 
in the field of 1 Einstein/IPC, 6 ROSAT/PSPC, and 4 ROSAT/HRI 
observations. We extracted the source events from 
the circular regions of 60'', 50'' and 20" radii 
centered on IKT~1 for the Einstein/IPC, ROSAT/PSPC 
and HRI observations, respectively. 
The  backgrounds are estimated using the same size 
regions in blank skies near the source. After the background 
subtraction and the vignetting correction, we estimated 
the source fluxes using the PIMMS software, and using 
the ASCA best-fit model \citep{yokogawa2003}.  
For comparison, the phase-averaged flux of 
the XMM-Newton observation was also estimated using the
same ASCA model. In Table \ref{tab:flux}, we summarize the
long-term X-ray fluxes (0.7$-$10.0~keV). 
We see that the XMM-Newton flux at MJD 51832 is 
higher than the Einstein, ROSAT and ASCA fluxes  
by more than $\sim 5$ times. Thus a large long-term flux 
variation is clearly found. It should be noted that, 
in the ASCA observation, IKT~1 exhibited a large flare 
of about 10$^{4}$~s duration \citep{yokogawa2003}.

\begin{table}
\begin{center}
\caption{Flux history of IKT~1}
 \begin{tabular}{lccc}   \hline
   Satellite/Detector & MJD & Count rate\footnotemark[$*$] & Flux\footnotemark[$\dagger$]
   \\ \hline
   Einstein/IPC      &   44189& $3.9\pm 0.8 $ & $5.2 \pm 1.1$\\
   ROSAT/PSPC  & 48538& $0\pm3$ &  $0\pm 5$  \\
\hspace{33pt}/PSPC  & 48728 & $11.0\pm0.8$& $17\pm 1$  \\
\hspace{33pt}/HRI   & 49098&$1.3\pm 0.6$ & $5\pm 2 $\\
\hspace{33pt}/PSPC  &49297&$ 2.0\pm 0.6$ & $2.9\pm 0.8$  \\
\hspace{33pt}/HRI   &49460&$ 0.9 \pm 1.0 $ & $ 4 \pm 4 $ \\
\hspace{33pt}/HRI   &49462&$ 1.2 \pm 1.4$ & $ 5 \pm 6 $ \\
\hspace{33pt}/HRI   &49639&$ -0.4 \pm 1.3$ & $ -2 \pm 5$  \\
   ASCA/GIS$^\ddagger$   &50765&\dotfill& $6.7$ \\
\hspace{27pt}/GIS$^\ddagger$   &51646&\dotfill & $14$\\
   XMM/PN   &51832& $234\pm3$& $29$\\ \hline
  \multicolumn{4}{@{}l@{}}{\hbox to 0pt{\parbox{85mm}{\footnotesize
      Notes. Uncertainties correspond to 1-$\sigma$ confidence regions.
      \par\noindent
      \footnotemark[$*$]Count rate ($\times 10^{-3}$~cnts~s$^{-1}$).
      \par\noindent
      \footnotemark[$\dagger$]X-ray flux 
      ($\times 10^{-13}$~ergs~cm$^{-2}$~s$^{-1}$) in the 0.7$-$10.0~keV band.
      \footnotemark[$^\ddagger$]Cited from \citet{yokogawa2003}.
     }\hss}}
\label{tab:flux}
 \end{tabular}
\end{center}
\end{table}

\section{Discussion}

\subsection{X-ray population}

The large long-term variability and the relatively low luminosity 
($\sim 10^{35-36}$~ergs~s$^{-1}$ 
from the ROSAT to XMM-Newton observations) 
suggest that IKT~1 is a wind-fed binary system. 
\citet{meyssonnier} detected an emission line star 
from a star cluster near IKT~1. 
Using the ROSAT observation, \citet{haberl2000} 
proposed that one of the stars in this cluster is an optical 
counterpart of IKT~1. 
We found only one cluster member, 
OGLE~004723.37-731226.9\footnote{The data available at 
http://bulge.astro.princeton.edu/\~{}ogle/}, is 
within the refined position error of IKT~1 by 
EPIC/PN observation (the radius of $\sim \timeform{0.7''}$), 
and hence this star must be an optical counterpart of IKT~1.
The long-term X-ray variation, the  pulsation 
of 263~s and  the power-law spectrum strongly support that 
IKT~1 is a Be/X-ray binary pulsar (Be/XBP). 

Since the space density of XBPs is extremely high in 
the south-west part of the SMC (see  Figure~\ref{fig:image}),
and  most of the XBPs are Be/XBPs or strong candidates 
(\cite{imanishi}, \cite{yokogawa2000b}b, \cite{yokogawa2000c}c),
this region must be an active star-formation site
of some  $\sim 10^{7}$~yrs ago. 
A H$_{\rm I}$ supergiant shell runs over this area 
(\cite{stanimirovic}) and the age of the shell is 
estimated to be about $10^{7}$~yrs. This supergiant shell, 
therefore, probably triggered the proposed star formation 
activity about $10^{7}$~yrs ago. 

\subsection{Origin of the Soft Component}

Although our results put IKT~1 to a strong candidate of Be/XBPs, 
the detailed X-ray features, the presence of the 
pulsating soft component and the absence of a pulsation in the 
power-law component, are rather peculiar compared
to the other Be/XBPs. The pulse profile of a flat top 
with a V-shape dip is also unique.

Among the about 30 XBPs (and candidates) in the Magellanic Clouds, 
the soft component has been observed from only a few sources: 
SMC~X-1 (\cite{marshall2003}; \cite{woo1995}), LMC~X-4 \citep{woo1996},
RX~J0059.2$-$7138 \citep{kohno2000}, EXO~05319$-$6609.2 
\citep{haberl2003}, and XTE~J0111.2$-$7317 (\cite{yokogawa2000a}a) 
(see also \cite{paul}, and references therein),
where the latter 3 are Be/XBPs. 
Thus IKT~1 may be the forth Be/XBP with a clear soft component.

For RX~J0059.2$-$7138, the soft component is modeled as 
a thin thermal plasma of $kT = 0.37$~keV and exhibits 
no pulsation \citep{kohno2000}. Thus the origin would be a 
largely extended plasma of a comparable size with the binary separation. 
EXO~053109$-$6609.2, on the other hand, shows a pulsation 
above 0.4~keV and both the soft and hard components may 
be pulsed \citep{haberl2003}. Thus the origin of the soft component 
must be comparable with or smaller than the size of a neutron star (NS). 
XTE~J0111.2$-$7317 has a peculiar spectrum of an inversely broken 
power-law model and shows pulsations both below and above 
the break energy (\cite{yokogawa2000a}a). 
Unlike these 3 sources, IKT~1 shows the X-ray pulsations only 
in the soft component. 

\citet{sasaki2003} reported that two XBPs 
in the SMC show pulsations mainly in the soft band.
AX~J0049.5$-$7323 (\cite{yokogawa2000c}c) 
and RX~J0101.3$-$7211 \citep{sasaki2001} also show 
pulsations mainly in the soft components.
The statistics of all the sources, however, were limited to
distinguish  whether or not the X-ray spectra are composed of pulsed soft 
and non-pulsed hard components. 
Accordingly, IKT~1 is the first object which exhibits a 
two-component spectrum with a pulsed soft component 
and a non-pulsed hard component. 
The soft component can be modeled with 
a BB radiation of 0.47~keV temperature. 
Using the X-ray luminosity of this BB 
in the high phase of the pulsation, the emission size is estimated to be  
$1.2\times 10^{2}$~km$^{2}$.  This size is about 10\% of 
the full surface of a NS, and hence the flux modulation
along the NS rotation would be expected. 
A possible origin of the soft X-rays might be 
an opaque shell at the magnetosphere, which was first proposed 
by \citet{mccray1976} for the pulsating soft emission from Her~X-1. 
Although the luminosity, hence the emission size of the soft emission 
from IKT~1 are significantly smaller than Her~X-1, this model 
is another possible origin of the pulsating soft X-ray emission.

\subsection {Origin of the Hard Component}

In general, XBPs exhibit a broken power-law spectrum in the hard 
component, which is likely coming from the small region near 
the pole, possibly the accretion column. 
The abnormal Thomson scattering in the strong magnetic 
field and the cyclotron resonance absorption/emission 
would be responsible for the spectral break. 
Since the hard component of IKT~1 has a typical broken power-law
spectrum, it is highly possible that these X-rays originate 
from the accretion column near at the magnetic pole.
A big puzzle, however, is that no coherent pulsation is found from
this power-law component.

Except pulsations, the power-law component is
highly variable with many flares with  about 4000-s quasi-periodicity. 
A plausible case is that  the 4000-s variation is due to  
quasi-periodic mass accretion. 
In fact, a long exposure (177~ks) observation 
of ASCA \citep{yokogawa2003} showed a large flare 
with the duration of about $10^4$~s. 
No hint of the 4000-s variations was, however, found in the flare. 
The fainter flux of IKT~1 and the smaller effective area 
of ASCA than that of XMM-Newton might have made 
it impossible to detect the 4000-s variations if any. 

Quasi-periodic oscillations (QPOs) at frequencies of 5$-$220~mHz 
have been observed from $\sim$10 XBPs to date (e.g., \cite{boroson},
and references therein). The canonical model of QPOs in XBPs 
is either a model of the beat-frequency between those of 
rotations of the NS ($\nu_{\rm ns}$) and the Kepler motion of 
the inner disk ($\nu_{\rm K}$) (a beat frequency model: BFM), 
or a Keplerian frequency model (KFM). 
In the BFM, the QPO frequency ($\nu_{\rm QPO}$) equals 
($\nu_{\rm K} - \nu_{\rm ns}$), where in the IKT~1 case 
$\nu_{\rm QPO} = 0.23$~mHz and $\nu_{\rm ns} = 3.82$~mHz. 
As a result, $\nu_{K}$ of IKT~1 must be about 4~mHz and the 
corresponding Keplerian radius is about $7 \times 10^{9}$~cm.
The disk matter at the inner Keplerian radius 
should be in pressure balance to the magnetospheric field.
Then assuming the radius and mass of 
NS to be $R_{0} = 15$~km, $m_{\rm x} = 1.4M_{\solar}$, 
and using the X-ray luminosity of 
$L_{\rm x} = 1.3 \times 10^{36}$~ergs~s$^{-1}$, 
we estimate the surface magnetic field strength of the NS 
to be $\sim 4 \times 10^{13}$~G (cf.\ \cite{davidson}). 
This value is beyond the range of normal NSs (or XBPs), 
and hence the BFM is unlikely.
In the KFM, the Keplerian orbital period
(frequency) of the inner disc is 
$\nu_{\rm K}$ = $\nu_{\rm QPO} = 0.2$~mHz, and hence
the radius is about $5 \times 10^{10}$~cm, which is
too far from the NS to produce the hard X-ray we observed.

In summary, IKT~1 exhibits  many unusual properties 
as an XBP, which must be a key for understanding the nature
and X-ray emission mechanisms of this source. 
For a more quantitative study of this peculiar pulsar, 
a very long exposure observation with high sensitivity 
instruments like XMM-Newton is highly anticipated.

\bigskip
The Einstein and ROSAT data are obtained through the High 
Energy Astrophysics Science Archive Research Center Online 
Service, provided by the NASA/Goddard Space Flight Center. 
M.U. and S.T. are supported by JSPS Research Fellowship for 
Young Scientists. This work is supported by a Grant-in-Aid for 
the 21 century COE, 
"Center for Diversity and Universality in Physics ".


\begin{thebibliography}{}

\bibitem[Anders \& Grevesse(1989)]{anders1989}
Anders, E., \& Grevesse, N.\ 1989, Geochimica et Cosmochimica Acta, 53, 197

\bibitem[Aschenbach \etal\ (2000)]{aschenbach}
Aschenbach, B., \etal\ 2000, Proc. SPIE, 4012, 731

\bibitem[Boroson et al.(2000)]{boroson} Boroson, B., O'Brien, 
K., Horne, K., Kallman, T., Still, M., Boyd, P.~T., Quaintrell, H., \& 
Vrtilek, S.~D.\ 2000, \apj, 545, 399 

\bibitem[Davidson \& Ostriker(1973)]{davidson} 
Davidson, K.~\& Ostriker, J.~P.\ 1973, \apj, 179, 585 

\bibitem[Haberl \& Sasaki(2000)]{haberl2000}
Haberl, F., \& Sasaki, M.\ 2000, \aap, 359, 573

\bibitem[Haberl, Dennerl, \& Pietsch(2003)]{haberl2003} Haberl, 
F., Dennerl, K., \& Pietsch, W.\ 2003, \aap, 406, 471 

\bibitem[Harries, Hilditch, \& Howarth(2003)]{harries} Harries, 
T.~J., Hilditch, R.~W., \& Howarth, I.~D.\ 2003, \mnras, 339, 157 

\bibitem[Imanishi et al.(1999)]{imanishi}Imanishi, K., 
Yokogawa, J., Tsujimoto, M., \& Koyama, K.\ 1999, \pasj, 51, L15

\bibitem[Inoue, Koyama \& Tanaka(1983)]{inoue}
Inoue, H., Koyama, K., \& Tanaka, Y.\ 1983, 
in IAU Symposium 101, Supernova Remnants and their X-Ray Emission, 
ed. J. Danziger \& P. Gorenstein (Dordrecht: D. Reidel Publishing Co.), 535 

\bibitem[Kohno, Yokogawa, \& Koyama(2000)]{kohno2000} Kohno, M., 
Yokogawa, J., \& Koyama, K.\ 2000, \pasj, 52, 299 

\bibitem[Marshall, Becker, \& White(1983)]{marshall2003} Marshall, 
F.~E., Becker, R.~H., \& White, N.~E.\ 1983, \apj, 266, 814 

\bibitem[McCray \& Lamb(1976)]{mccray1976} McCray, R.~\& Lamb, 
F.~K.\ 1976, \apjl, 204, L115 

\bibitem[Meyssonnier \& Azzopardi(1993)]{meyssonnier}
Meyssonnier, N., \& Azzopardi, M.\ 1993, \aaps, 102, 451

\bibitem[Morrison \& McCammon(1983)]{morrison1983} Morrison, R.~\& 
McCammon, D.\ 1983, \apj, 270, 119 

\bibitem[Paul et al.(2002)]{paul} Paul, B., Nagase, F., 
Endo, T., Dotani, T., Yokogawa, J., \& Nishiuchi, M.\ 2002, \apj, 579, 411 

\bibitem[Sasaki, Haberl, Keller, \& Pietsch(2001)]{sasaki2001} 
Sasaki, M., Haberl, F., Keller, S., \& Pietsch, W.\ 2001, \aap, 369, L29 

\bibitem[Sasaki, Pietsch, \& Haberl(2003)]{sasaki2003}
Sasaki, M., Pietsch, W., \& Haberl, F.\ 2003, \aap, 403, 901

\bibitem[Stanimirovi\'{c} \etal\ (1999)]{stanimirovic}
Stanimirovi\'{c}, S., Staveley-Smith, L., Dickey, J.~M., Sault, R.~J., 
Snowden, S.~L.\ 1999, \mnras, 302, 417

\bibitem[Str$\ddot{\rm u}$der et al.(2001)]{struder}
Str$\ddot{\rm u}$der, L., \etal\ 2001, \aap, 365, L18

\bibitem[Turner et al.(2001)]{turner}
Turner, M.~J.~L.,  \etal\ 2001, \aap, 365, L27

\bibitem[Wang \& Wu(1992)]{wang}Wang, Q., \& Wu, X.\ 1992, \apjs, 78, 391

\bibitem[Woo et al.(1995)]{woo1995} Woo, J.~W., Clark, G.~W., 
Blondin, J.~M., Kallman, T.~R., \& Nagase, F.\ 1995, \apj, 445, 896 

\bibitem[Woo et al.(1996)]{woo1996} Woo, J.~W., Clark, G.~W., 
Levine, A.~M., Corbet, R.~H.~D., \& Nagase, F.\ 1996, \apj, 467, 811 

\bibitem[Yokogawa et al.(2000a)]{yokogawa2000a} Yokogawa, J., 
Paul, B., Ozaki, M., Nagase, F., Chakrabarty, D., \& Takeshima, T.\ 
2000a, \apj, 539, 191 

\bibitem[Yokogawa et al.(2000b)]{yokogawa2000b}Yokogawa, J.,
Torii, K., Imanishi, K., \& Koyama, K.\ 2000b, \pasj, 52, L37

\bibitem[Yokogawa et al.(2000c)]{yokogawa2000c}Yokogawa, J.,
Imanishi, K., Ueno, M., \& Koyama, K.\ 2000c, \pasj, 52, L73

\bibitem[Yokogawa et al.(2003)]{yokogawa2003} Yokogawa, J., 
Imanishi, K., Tsujimoto, M., Koyama, K., \& Nishiuchi, M.\ 2003, \pasj, 55, 
161

\end{thebibliography}
\end{document}